\documentstyle[aps]{revtex}

\twocolumn

\begin{document}
\title{Quantum tomography as normalization of incompatible observations }
\author{Z. Hradil,\cite{optika} J. Summhammer, H. Rauch}
\address{Atominstitut der \"{O}sterreichischen Universit\"{a}ten,\\
Stadionallee 2,\\
A--1020 Wien, Austria}
\date{\today}
\maketitle

\begin{abstract}
 Quantum states are successfully reconstructed
using the maximum likelihood estimation on the subspace where the
measured  projectors
reproduce the identity operator.
Reconstruction corresponds to    normalization
of incompatible observations.
 The proposed approach handles the noisy
data corresponding to realistic incomplete observation with finite
resolution.
\end{abstract}


Quantum theory handles observable events on the most fundamental level
currently available predicting the statistics of quantum phenomena. This
randomness is hidden in the quantum state. Although the history of state
reconstruction may be traced back to the early days of quantum mechanics to
the Pauli problem \cite{St}, till quantum optics opened the new era.
Theoretical prediction of Vogel and Risken \cite{VR} was closely followed by
the experimental realization of the suggested algorithm by Smithey et. al.
\cite{SBRF93}. Since that time many improvements and new techniques have
been proposed, an overview can be found in Ref. \cite{Ulf}. Reconstruction
of quantum states is considered as standard technique used in various
branches of contemporary physics \cite{endo}. Unfortunately, there are
several flaws in the approach. The available measurement is always limited
as far as the amount and accuracy of data is concerned. Standard methods are
designed for analysis of sharp, complete and noiseless observations. When
applied on the realistic data, serious problems with positivity of
reconstructed density matrix appear. Though these techniques may give a
rough picture of the state, they cannot provide a full quantum description.

This may be  accomplished using the maximum likelihood (MaxLik) estimation
\cite{Hradil1}. The question of deterministic schemes: ``What quantum state
is determined by that measurement?'' is replaced by the formulation
consistent with quantum theory: ``What quantum states seem to be most likely
for that measurement?'' Quantum theory predicts the statistics provided that
quantum state is known. The fundamental result of this Letter is the
statistical inversion of this quantum postulate predicting the quantum state
provided that results of the measurement are known. General theory is
formulated for the case of nonorthogonal measurement of
incompatible observables. Physically it corresponds to synthesis of
various measurements done under different experimental
conditions,
always performed on the identically prepared system. It might be
subsequent recording of an unknown spin of the neutron using different
settings of the Stern Gerlach apparatus, or the recording of the quadrature
operator for light in the rotated frame in quantum tomography.
 The formulation provided here will
focus the general aspects illustrating the novel approach on the example of
quantum tomography.

In the following the states $|y_i\rangle $ will denote general nonorthogonal
states enumerated for concreteness but without loss of generality by a
discrete quantum number. Assume that the quantum measurement has been done
$n $ times yielding the
relative frequencies of the event represented by the
projector $|y_i\rangle $ as $f_{i}>0.$ The states are assumed to be
nonorthogonal and linearly independent. Hence the correlation matrix is
invertible and Hermitian $C_{ij}=C_{ji}^{*}=\langle y_i|y_j\rangle .$ {\em %
(i)} The measurement is sharp provided that it may be described by a
projector into the pure state $|y_i\rangle \langle y_i|.$ On the contrary,
projectors corresponding to unsharp measurement are given by the probability
operator measure \cite{Hel76} representing a superposition in bins $D_{i}$
of indistinguishable states $\hat{\Pi}_{i}=\sum_{j \in D_{i}}|y_j\rangle
\langle y_j|.$ {\em (ii)} Measurement is complete if all the projectors
corresponding to the counted data yield the decomposition of unity operator,
$\sum_{i,all}|y_i\rangle \langle y_i|=\hat{1}.$ The states are assumed to be
normalized with respect to this full
completeness relation but not to the scalar
product as usually. Provided that the counted data do not exhaust all the
values, the measurement is incomplete. In the notation used here, the
decompositions are assumed to be incomplete unless defined otherwise. {\em %
(iii)} The data are noise free provided that the counted frequencies
coincide exactly with the prediction of quantum theory
\begin{equation}
f_i= \rho_{ii} \equiv \langle y_i| \hat \rho |y_i \rangle  \label{det}
\end{equation}
for some density matrix $\hat \rho $ and data are noisy otherwise. Sharp,
complete and noise free measurement is assumed implicitly in the standard
description of quantum tomography based on the direct inversion of (\ref{det}%
). When this ideal conditions are not met, the algorithm does not guarantee
the positive definiteness of the reconstructed ``density matrix".

Realistic data can never provide complete information about quantum system
with infinite degrees of freedom. This may be demonstrated on the simplified
example of reconstruction of diagonal elements of density matrix via photon
counting with ideal detector. Suppose $n$ times repeated counting, always
with the zero registered photoelectrons. The ``standard'' prediction of
quantum state reads $\hat \rho =|0\rangle \langle 0|,$ where $|0\rangle $
denotes the vacuum state. Nevertheless this interpretation may be always
spoiled by the classical noise represented by a projector into the strong
coherent state $\hat{{\cal N}}_{\epsilon }= |\frac{\alpha }{\sqrt{\epsilon }}%
\rangle _{coh}\langle \frac{ \alpha }{\sqrt{\epsilon }}| $ appearing with
the negligible probability $\epsilon .$ Obviously, the state $\hat{\rho}%
_{\epsilon }=(1-\epsilon )\hat{\rho} + \epsilon \hat{{\cal N}}_{\epsilon }$
cannot be distinguished from the standard one for sufficiently small $%
\epsilon <1/n.$ This may appear as crucial for some observations. For
example, the average numbers of particles differ significantly for both the
states. The state of the system is well defined as the ``standard'' density
matrix on the scanned part of the Hilbert space spanned by the actually
measured values. The behavior of the wave function in the complementary
world is t unknown. If some future detection of a quantum
variable will depend significantly just on this ``unobserved'' part, the
prediction should be uncertain. This seems to be evident in the case of
orthogonal measurements but as will be seen, rather nontrivial in
case of nonorthogonal measurements.

Quantum state attached to the data will be searched as the density matrix $%
\hat{\rho}$ which maximizes the likelihood functional \cite{Hradil1}
\begin{equation}
{\cal L}(\hat{\rho})=\prod_{i}\langle y_{i}|\hat{\rho}|y_{i}\rangle
^{nf_{i}}.  \label{lik}
\end{equation}
Assume the diagonal representation of a density matrix as
\begin{equation}
{\hat{\rho}}=\sum_{k}r_{k}|\phi _{k}\rangle \langle \phi _{k}|.
\label{density}
\end{equation}
The existence of parameters $r_{k}\geq 0,\sum_{k}r_{k}=1$ and an orthogonal
basis $|\phi _{k}\rangle $ is guaranteed by quantum theory. Normalized
extremum states satisfy the relation
\[
\frac{\partial }{\partial \langle \phi _{k}|}\biggl[\frac{1}{n}\ln {\cal L}%
-\Lambda {\rm Tr}(\rho )\biggr]=0,
\]
$\Lambda $ being a Lagrange multiplier. This reads the system of coupled
equations
\begin{equation}
\hat{R}|\phi _{k}\rangle =|\phi _{k}\rangle ,  \label{vysledek2}
\end{equation}
\[
\hat{R}=\sum_{i}\frac{f_{i}}{\rho _{ii}}|y_{i}\rangle \langle
y_{i}|,\;\;\;\rho _{ii}=\sum_{k}r_{k}|\langle \phi _{k}|y_{i}\rangle |^{2}.
\]
In the derivation, the condition of normalization ${\rm Tr}\hat{\rho}=1$ has
been used. Relation (\ref{vysledek2}) provides the statistical inversion of
quantum postulate (\ref{det}) as the nonlinear equation for density matrix
\begin{equation}
\hat{R}(\hat{\rho})\hat{\rho}=\hat{\rho}.  \label{vysledek}
\end{equation}
Reconstruction is done in the subspace where operator $\hat{R}$ represents
the identity operator. The equations for matrix elements $\rho _{ij}$ read
\begin{equation}
f_{i}\rho _{ij}=\rho _{ii}\sum_{k}C_{ik}^{-1}\rho _{kj},  \label{mat1}
\end{equation}
$C^{-1}$ being the inversion matrix to $C_{ij}=\langle y_{i}|y_{j}\rangle .$
Notice that the diagonal elements
 are instead of (\ref{det}) fulfilling the relation
\begin{equation}
C_{ii}^{-1}\rho _{ii}+\sum_{k\neq i}C_{ik}^{-1}\rho _{ki}=f_{i}.
\label{mat2}
\end{equation}
Though the relation (\ref{det}) is linear with respect to the density
matrix, the inversion represented by (\ref{vysledek}) or by (\ref{mat1}) is
not. The reasons  are fundamental: Elements of density matrix are not
independent, but characterize a  quantum state. These
quantum correlations are neglected when the inversion is done regardless on
the positive definiteness. To find solutions of nonlinear operator equation (%
\ref{vysledek}) is a peculiar problem. It may be approached iteratively
provided that necessary conditions for convergence are fulfilled. This
questions will be addressed separately elsewhere. Formulation simplifies
considerably provided that the projectors in operator $\hat{R}$ commute. The
density matrix is diagonal in this common basis. Reconstruction of diagonal
elements of density matrix using the homodyne detection with random phase
represents an explicit example \cite{Banaszek} and iterative algorithm is
very effective here. Solution need not be unique depending in general on the
starting point of iterations. Consequently, MaxLik estimation provides a
family of extremum states not distinguished by the given measurement.
Averaging over this family enhances the uncertainty of state prediction
confirming the conjecture formulated in \cite{Hradil1}. The analysis of
realistic measurement supporting this interpretation is given in \cite
{Hradil2}.

Relations (\ref{lik}) and (\ref{vysledek2}) may show, how closely the
given state approaches the extremum one. The relative entropy
(normalized log likelihood)
\begin{equation}
K(\rho /f)=-\sum_{i}f_{i}\ln \frac{\rho _{ii}}{f_{i}}\geq 0
\label{vysledek3}
\end{equation}
provides the difference between absolute minimum and estimated result. Its
value may be expressed in $\%$ of the entropy $S=-\sum_{i}f_{i}\ln f_{i}.$
Similarly, the ``experimentally achieved resolution of identity'' $\hat{R}$
could always be compared with the identity operator.

The MaxLik quantum  state reconstruction possess very clear
geometrical interpretation as
 {\em  normalization of incompatible
observations}. Indeed, the rays in Hilbert space are given up to the
multiplicative factors. The renormalized projectors
 $|y_{i}\rangle \rightarrow
|y_{i}^{\prime }\rangle =\sqrt{f_{i}/\rho _{ii}}|y_{i}\rangle $
fulfill  the
relation analogous to (\ref{det}) as
\begin{equation}
\label{renorm}
\langle
y_i^{\prime}| \hat \rho | y_i^{\prime}\rangle = f_i.
\end{equation}
Moreover the  operator $\hat R $
characterizes the overlapping of rays in analogy with tomography in
medicine. The X--rays overlap in the scanned region illuminating larger part
of the body. In the language of quantum theory the linear envelope of the
detected projectors represents an analogy of the whole irradiated space $%
{\cal U}=\{|y_{i}^{\prime}\rangle \},$
a subspace of the full Hilbert space ${\cal H}.$
 Denote formally the orthogonal subspace of overlapping of projectors as $%
{\cal O},$ $|x_{k}\rangle $ being its orthogonal basis. Each projector can
be then decomposed as
$
|y_{i}^{\prime}\rangle =\sum_{k}\langle x_{k}|y_{i}^{\prime}
\rangle |x_{k}\rangle
+|Z_{i}\rangle ,
$
where $|Z_{i}\rangle $ are orthogonal to subspace ${\cal O},$ $\langle
x_{k}|Z_{i}\rangle =0.$ The sum of scalar products $Z=\sum_{i}\langle
Z_{i}|Z_{i}\rangle $ characterizes the part of the projectors outside the
orthogonal subspace ${\cal O}.$ The basis $|x_{k}\rangle $ will be chosen in
order to minimize $Z$ under the condition of normalization. The
optimum  basis in ${\cal O}$ is spanned by the eigenstates diagonalizing the
sum of projectors achieved by the realistic measurement
\begin{equation}
\biggl[\sum_{i}|y_{i}^{\prime}\rangle \langle y_{i}^{\prime}
|\biggr]|x_{k}\rangle =\lambda
_{k}|x_{k}\rangle .  \label{vysledek1}
\end{equation}
Hence
MaxLik reconstruction  (\ref{vysledek2}) may be easily
interpreted in the language of  quantum (\ref{renorm}) and geometrical
(\ref{vysledek1})  considerations.
The operator $ \hat{R} =\sum_{i}|y_{i}^{\prime }\rangle
\langle y_{i}^{\prime }|  $ characterizes the overlapping of projectors.
The reconstruction should be done in a
subspace ${\cal O}_{rec}$ spanned by the eigenstates $|x_{k}\rangle $ with
the degenerate eigenvalue $\lambda _{k}=1.$
 The density matrix is
spanned in the subspace  where $\hat{R} $ equals identity operator.
The quantum postulate (\ref{det}) predicting the statistics of
the outcome provided that quantum state is known should be
modified to formally  analogous relation (\ref{renorm}) provided that an
inversion problem has to be solved. However, the later problem
is nonlinear since the normalization of projectors depends
on the state itself.
Any  reconstruction beyond the
subspace ${\cal O}_{rec}$ is rather a random guess since not enough
information is available. In medicine, it would correspond to the
``observation'' of the head while stomach has been scanned, for example. The
subspaces are related as ${\cal O}_{rec}\subset {\cal O}\subset {\cal U}%
\subset {\cal H},$ but may coincide in some special cases with low degrees
of freedom, as for example in the case of spin systems. Since only the
decomposition of identity matters, the formulation is common for both the
sharp and unsharp observations. The MaxLik estimations simplifies
considerably, provided that measured projectors $\hat \Pi_i$ commute.
Observations are compatible and renormalization is not necessary. The
subspace ${\cal O} $ may be approximately characterized in the common basis $%
|\xi\rangle $ by diagonal elements of operator $\hat R $ for $\rho_{ii}= f_i$
\begin{equation}
R(\xi )=\langle \xi |\biggl[ \sum_i \hat \Pi_i \biggr] |\xi \rangle.
\label{fidel}
\end{equation}
The function $R(\xi)$ plays the role of Optical Transfer Function in Fourier
optics characterizing the fidelity of the information about variable $\xi $
contained in the measured data.

The theory is illustrated on numerical simulations of the quantum tomography
experiments of the type \cite{SBRF93,Schiller}. For quantum tomography, the
projectors are given by rotated quadrature states $|y_{ij}\rangle =1/\sqrt{%
\pi }|x_{i},\theta _{j}\rangle $ corresponding to the center of the
coordinate--phase bins. For random--phase homodyning the commuting
projectors are $\hat{\Pi}_{i}=1/(2\pi )\int_{0}^{2\pi }d\theta |x_{i},\theta
\rangle \langle x_{i},\theta |.$ Data depicted in the Fig.~\ref{fig1}
correspond to the tomography of squeezed vacuum state--an eigenstate of the
operator $\hat{b}=\hat{a}\cosh r+e^{i\varphi }\sinh r\hat{a}^{\dagger }$ for
$r=1,\varphi =\pi /2.$ Scanning has been done at 12 phase cuts using 600
records at each phase position. The scanned intervals $(-7,7)$ in each cut
are divided into 100 bins. An ideal detection $(\eta =1)$ is assumed.
Solution of the nonlinear eq. (\ref{vysledek}) represents the key point of
the reconstruction. An iteration procedure has been already applied to
random--phase homodyning \cite{Banaszek,Hradil2}. General iteration
procedure in tomography will be dealt with elsewhere. For the purpose of
illustration the likelihood is maximized numerically using MATLAB for pure
states. Estimated complex amplitudes $\Psi _{n}$ in number--state basis
(hollow bars) are compared with the true amplitudes (full bars) in the left
panels of the Fig. \ref{fig2}. The typical oscillating nature is obvious
here. The hollow bars in right panels of the Fig. \ref{fig2} show the
diagonal elements of the operator $\hat{R}$ in $28$ dimensional (upper
panel) and in $25$ dimensional subspaces. The data are insufficient for MaxLik
reconstruction in the former case  since the identity operator is
not recovered, but are sufficient in the later case.
For comparison, the full bars show the decomposition of identity
for random--phase homodyning \cite{indirect,Banaszek}, where the
reconstruction can be done in about $10$ dimensional subspace only. The
relative entropy of the true state and generated data is about $K(\rho
/f)=0.95\%$ of the entropy of measured data $S=5.89.$ The MaxLik fitting
provides the value $K(\rho /f)=0.81\%$ of the entropy. For comparison, a
random guess is characterized by the value of several tens of $\%$ of the
entropy $S.$ Finally, the Fig. \ref{fig3} shows the proper normalization of
the projectors $\sqrt{f(x_{i},\theta _{j})/\rho (x_{i},\theta _{j})},$ by
which the completeness relation is fulfilled. Not registered projectors are
missing here and events registered with low relative frequencies are
renormalized.

MaxLik procedure provides an effective method for statistical inverting of
quantum postulate (\ref{det}). It fits the data better than deterministic
schemes, but may provide a family of indistinguishable states as the result.
MaxLik algorithm is nonlinear and standard error analysis cannot be applied.
Particularly, reconstruction may be accomplished on the subspace where
measured projectors provide the resolution of identity. These issues should
be taken into account since the state reconstruction has became to play an
important role in many sophisticated detection techniques.

Many discussions with A. Zeilinger, S. Weigert, S. Schiller, T. Opatrn\'{y},
J. Pe\v{r}ina. M. Du\v{s}ek and R. My\v{s}ka are appreciated. This work was
supported by TMR Network ERB FMRXCT 96-0057 ``Perfect Crystal Neutron
Optics'' of the European Union and by the grant of Czech Ministry of
Education VS 96028.

\begin{figure}[tbp]
\caption{Histograms of homodyne detection for squeezed vacuum.}
\label{fig1}
\end{figure}

\begin{figure}[tbp]
\caption{ Left panels show the real and imaginary parts of the true (full)
and estimated (hollow) amplitudes of pure state in number state basis. Right
panels show the diagonal elements of decomposition of operator $\hat R$ for
random-phase homodyne detection (full) and homodyne tomography (hollow). For
homodyne tomography, data are insufficient for reconstruction in $28$
dimensional subspace (upper panel), but are sufficient to recover the
identity in 25 dimensional subspace (lower panel). }
\label{fig2}
\end{figure}

\begin{figure}[tbp]
\caption{Renormalization of projectors in the nonorthogonal rotated
quadrature state basis corresponding to successful reconstruction. }
\label{fig3}
\end{figure}

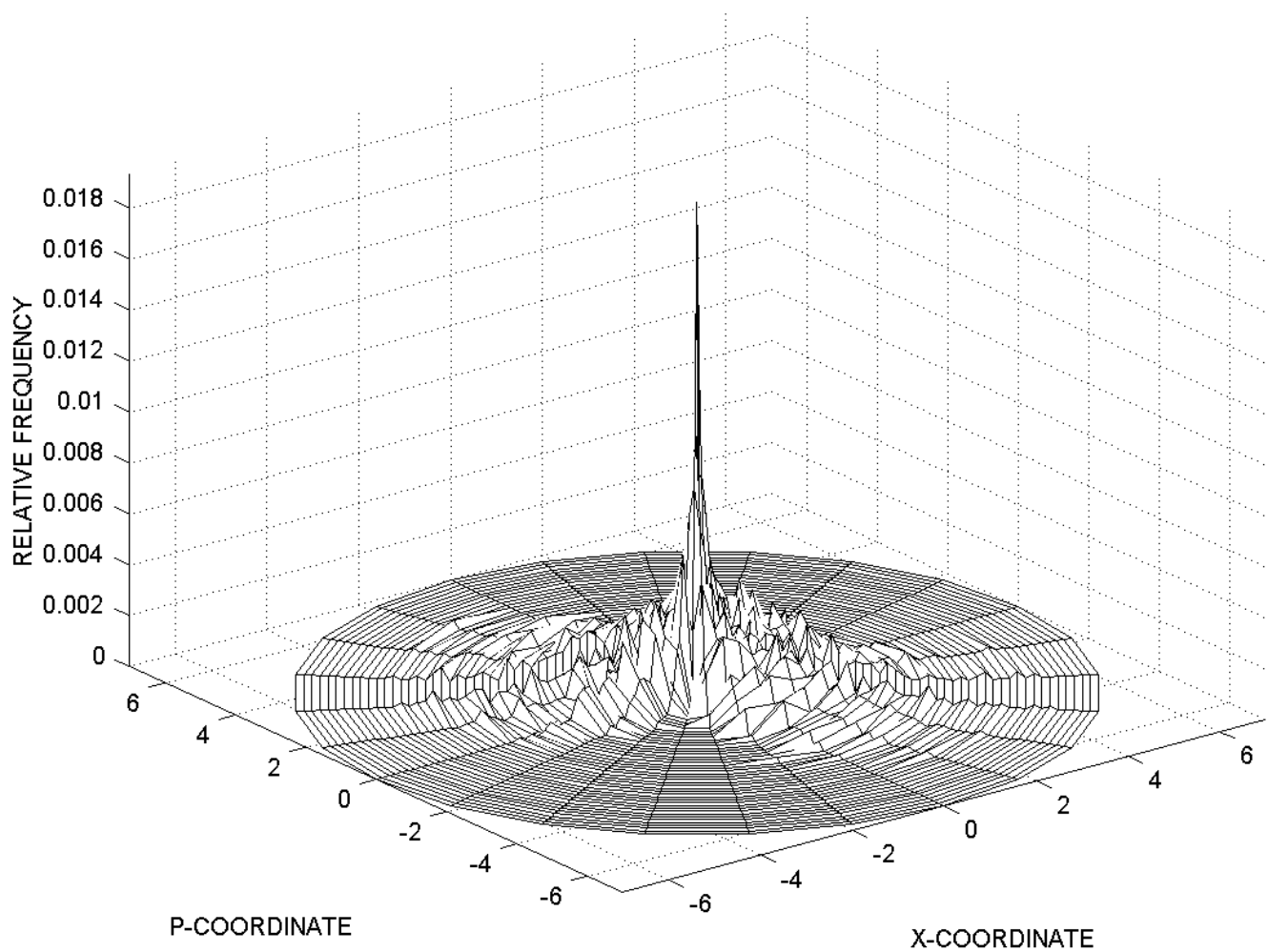
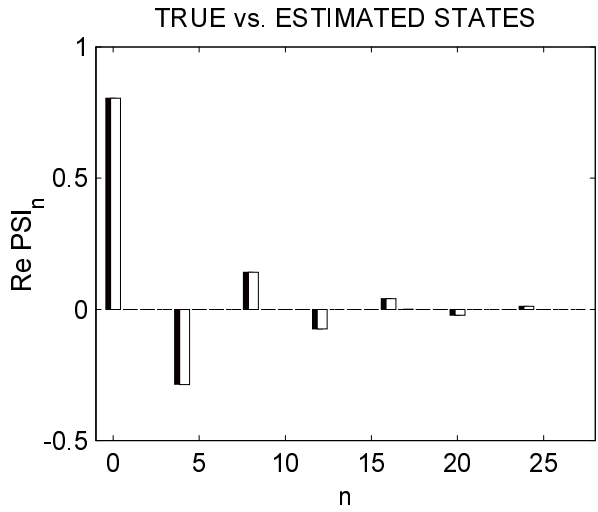
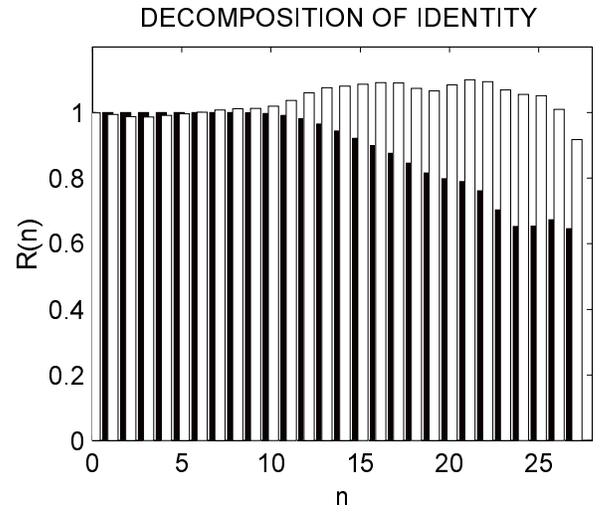
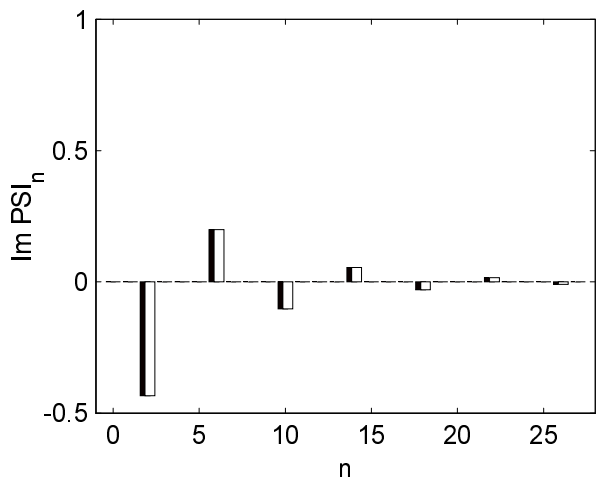
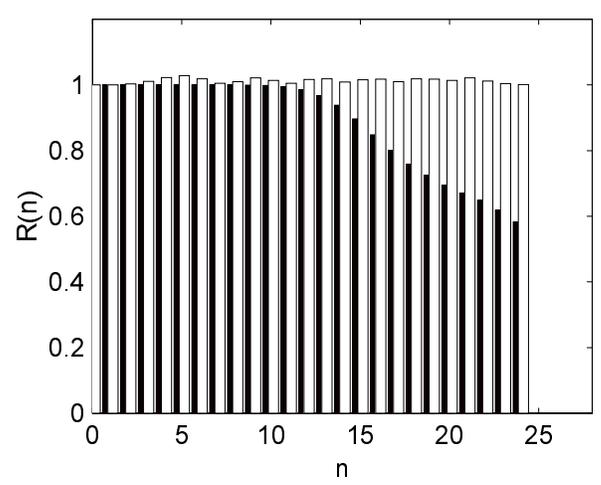
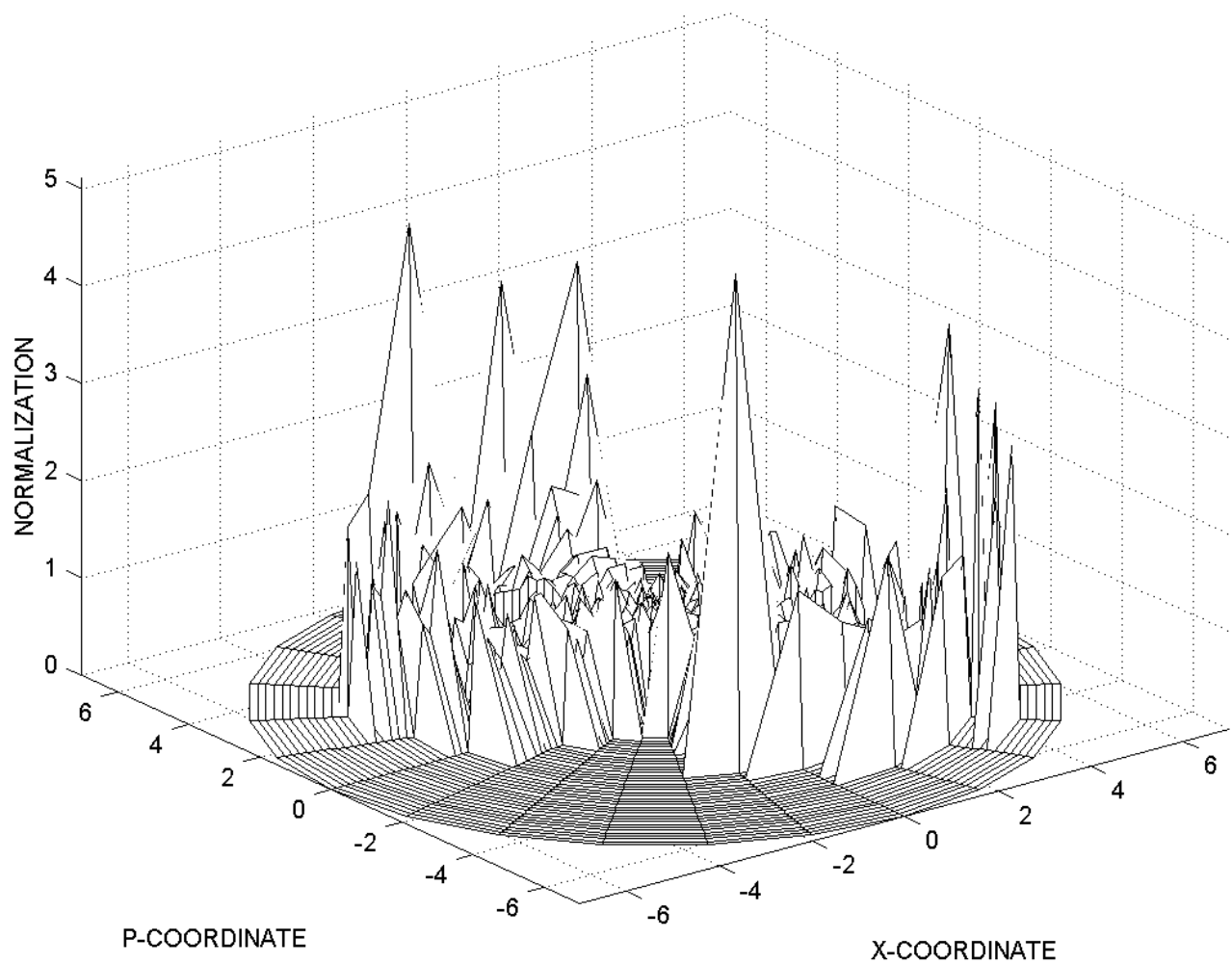
\end{document}